# Analysis of the rich optical iron-line spectrum of the x-ray variable I Zw 1 AGN 1H0707–495

**H Winkler, B Paul**

Department of Physics, University of Johannesburg, PO Box 524, 2006 Auckland Park, Johannesburg, South Africa

E-mail: hwinkler@uj.ac.za

**Abstract**. Thirty years ago the optical counterpart of the x-ray source 1H0707-495 was discovered to be a 15th magnitude broad-line Seyfert galaxy with a rich Fe II emission line spectrum typical of the AGN subclass sometimes referred to as the I Zw 1 objects after their progenitor. This object became the subject of much interest and investigation just over five years ago when it was shown to have undergone dramatic x-ray luminosity variations. This paper presents an extensive series of medium resolution spectra recorded at the 1.9 m telescope at Sutherland in January 2016. Through co-adding the spectra, we are able to achieve a signal-to-noise hitherto not achieved for this object, allowing us to resolve individual Fe II lines and measure their relative strengths and profiles to a degree of accuracy not previously available for this AGN. We provide possible physical interpretations of our measurements and investigate links between the spectral evidence collected in this study and the known x-ray behaviour.

## 1. Introduction

The advent of x-ray astronomy and subsequent systematic searches of the optical counterparts of x-ray-bright sources led to the identification of a significant number of these with active galactic nuclei (AGN). Optical spectroscopy revealed the x-ray source 1H0707–495 to be linked to a galaxy of magnitude $V = 15.7$ mag at redshift $z = 0.041$ with a rich emission line spectrum with broadened features indicative of Seyfert type 1 class [1]. The spectrum distinguished itself from that of other AGN identified in the discovery paper through its uncharacteristically strong Fe II emission spectrum, no obvious narrow forbidden lines and the fact that the Balmer lines were not as broad. This type of spectrum has subsequently been recognized as a Seyfert/AGN subclass of its own, referred to as narrow line Seyfert 1 (NLS1) [2], and, when the iron-line spectrum is exceptionally prominent, as I Zw 1 type object after the sub-class prototype of that name.

As with other AGN, the objects associated with the I Zw 1 sub-class are believed to consist of a central massive black hole surrounded by an accretion disk illuminating nearby fast-moving gas. However the exact mechanisms generating their characteristic spectra are not properly understood. It is not even clear why the spectra of some broad-line Seyfert galaxies display no iron emission lines whatsoever, while there are other AGN such as Markarian 231 where the iron spectrum completely dominate the blue part of the spectrum (see e.g. [3]).

1H0707–495 has in the more recent past been frequently observed in other wavelength regimes, particularly in x-rays [4], where strong variability has been detected. This contrasts with optical luminosity variations that are quite moderate by comparison (e.g. [5]).

## 2. Observations and spectral calibration

Optical spectroscopic observations of 1H0707–495 were carried out with the 1.9 m telescope at the Sutherland station of the South African Astronomical Observatory in January 2016. We used the medium resolution reflection grating with 600 lines/mm corresponding to a dispersion of 100 Å/mm, and set the slit width to 150 μm, translating to 0.9 arcsec in the field of view. The grating angle was adjusted to obtain a wavelength range $\lambda\lambda$ 3600-6400 Å on the detector.

Two sets of spectra were collected. The first set, obtained during the night 13-14 January under dark sky conditions, consisted of four exposures totaling 3600 s. The second set (three exposures totaling 3000 s) was recorded on the night 19-20 January with a bright moon up (lunar phase = 0.81). The CCD images obtained were processed using standard procedures: i) the frame bias was subtracted; ii) intrinsic pixel sensitivity differences and throughput variations along the slit were corrected using flatfield images obtained from illumination by a smooth-spectrum lamp; iii) cosmic ray blemishes on the images were removed by manually identifying affected pixels and smoothing these using neighbouring pixels.

The wavelength calibration was effected by means of argon calibration spectra recorded before and after each spectrum of 1H0707–495. The resultant plots of wavelength versus pixel position were fitted by means of third-order polynomials, which were then applied to the galaxy spectrum. The sky background was determined by averaging the recorded (and similarly wavelength calibrated) spectrum in two strips 15 CCD channels on either side of the galaxy – no evidence of an extended galaxy halo could be detected in these strips. The sky spectrum was then subtracted from the preliminary object spectrum to obtain the non-contaminated spectrum of the AGN. The noise levels of the three spectra of 19 Jan are significantly higher than for those of 13 Jan due to the bright moonlit sky later in the week.

Conditions were not photometric on both nights, and the seeing also not stable, meaning that absolute spectrophotometric calibration was not possible. Nonetheless, spectra of the spectrophotometric standard star LTT 2415 [6] were recorded and, together with the standard Sutherland atmospheric extinction correction, used to determine the relative spectral flux calibration, so that it becomes possible to determine emission line ratios (but not absolute line fluxes). No corrections were applied for telluric absorption as the significant telluric lines all fall outside the observed wavelength range.

The seven individual spectra for 1H0707–495 were finally combined to obtain the high signal-to-noise spectrum illustrated in figure 1.

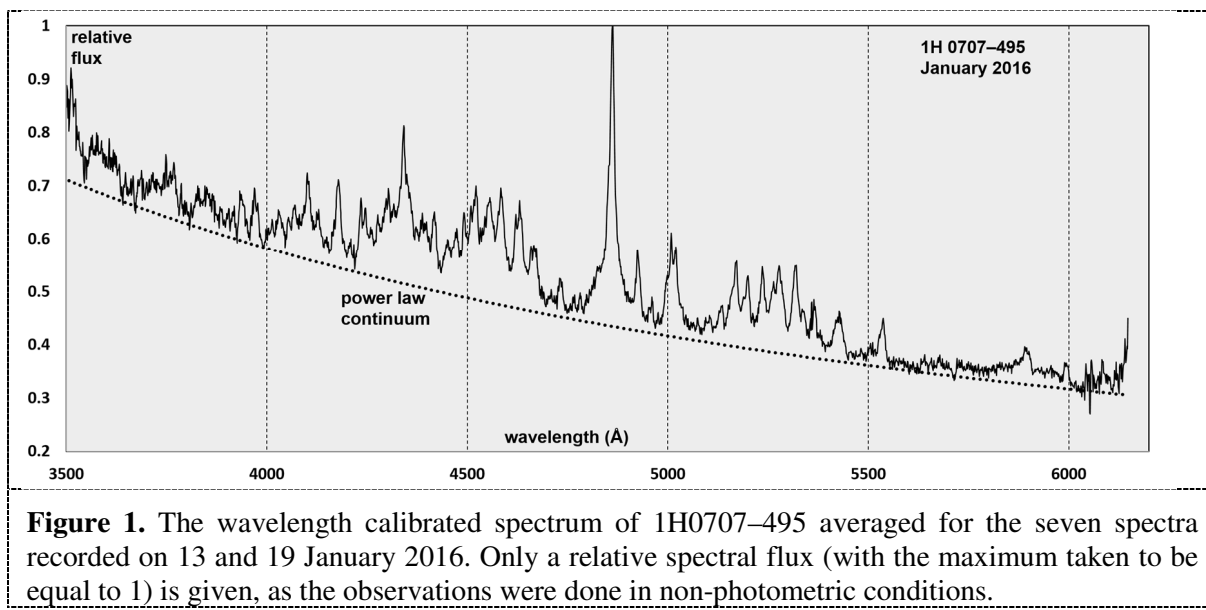

**Figure 1.** The wavelength calibrated spectrum of 1H0707–495 averaged for the seven spectra recorded on 13 and 19 January 2016. Only a relative spectral flux (with the maximum taken to be equal to 1) is given, as the observations were done in non-photometric conditions.

## 3. Spectral analysis
### 3.1. General spectral characteristics
The January 2016 spectrum of 1H0707–495 still closely resembles the discovery spectrum of Remillard et al. [1], confirming that the I Zw 1 classification still applies. While a fraction of the photons recorded in the spectrum originate from integrated host galaxy starlight generated far from the nucleus, that component is not distinguishable in figure 1. There is also no sign of interstellar absorption features at $\lambda\lambda$ 3934/67 Å (Ca II) and $\lambda\lambda$ 5889/96 Å (Na I) that can be conspicuous in similar iron-rich AGN spectra.

The overall shape of the continuum is well fitted by a power law spectrum $F_\lambda \propto \lambda^\gamma$ with a spectral index $\gamma = -1.5$ (also shown in figure 1). This agrees closely with the estimate from UV-optical spectra from previous studies [7].

### 3.2. The Fe II spectrum
We then introduce the Fe II spectral templates computed by Véron-Cetty et al. [8] and Kovacevic et al. [9], and compare this with the observed spectrum (with the power law subtracted) in figures 2 and 3. The former template was determined empirically from a spectrum of I Zw 1, while the latter is based on theoretical calculations of iron ion transition probabilities, with the full width at half maximum (*FWHM*) adjusted to 800 km s$^{-1}$.

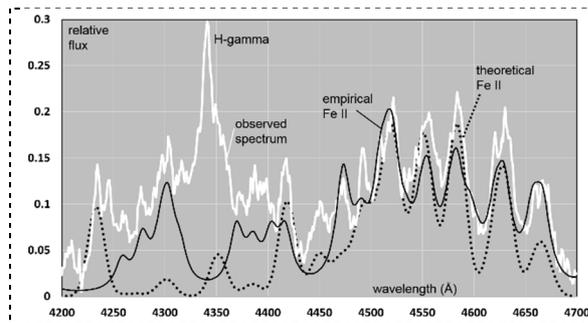 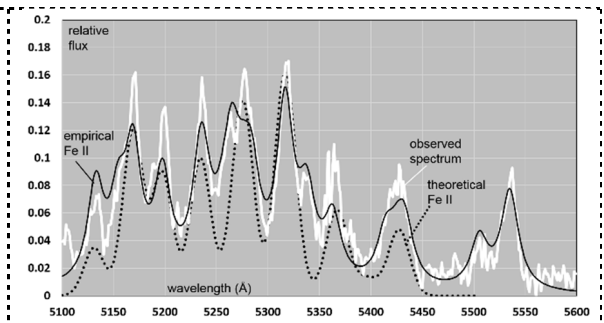

**Figure 2.** Fit of the observed Fe II emission bands near 4500 Å.

**Figure 3.** Fit of the observed Fe II emission bands near 5300 Å.

While most of the peaks in the templates match actually measured peaks, neither of the templates fit the observed spectra satisfactorily.

### 3.3. The H-beta emission profile
The H-beta emission line has been fitted by the superposition of two commonly applied profile functions (see figure 4). The first of these is a comparatively narrow Gaussian component with a *FWHM* corresponding to 800 km s$^{-1}$, which is identical to the width most suited to fitting the Fe II emission line spectrum.

The broader of the H-beta components was best fitted by a Lorentzian profile shifted bluewards from the other component by 120 km s$^{-1}$, and with a *FWHM* of 2600 km s$^{-1}$. While the Gaussian component has a greater peak height, the total flux ratio of the Lorentzian H-beta fraction to the Gaussian fraction is 3.7. There is an additional bump in the red wing of the line (near 4850 Å), which may be due to a further component in H-beta or another feature related to the iron line spectrum.

### 3.4. Forbidden line spectrum
Figure 5 displays the part of the spectrum containing the characteristic AGN nebular lines [O III] 4959 Å and 5007 Å. While these clearly are exceptionally weak, small peaks corresponding to these features are clearly visible with wavelengths almost consistent with redshift of the H-beta peak.

These peaks have been fitted with Gaussian profiles, maintaining their theoretically known peak ratio of 1:3. The width thus obtained for these forbidden lines only correspond to *FWHM* ~ 400 km s$^{-1}$. We measure the flux ratio of the (total) H-beta line to the [O III] 5007 Å line, a commonly used indicator of the nature of the AGN activity [10], to be 42:1. This is exceptionally high (by comparison, the ratio rarely exceeds 10:1 in type 1 Seyfert galaxies), and points to possible fundamental differences in the nuclear structure of 1H0707–495.

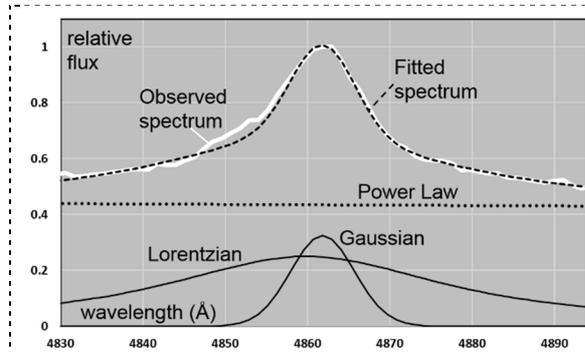

**Figure 4.** The hydrogen 4861 Å (Hβ) emission line fitted as a superposition of Gaussian components.

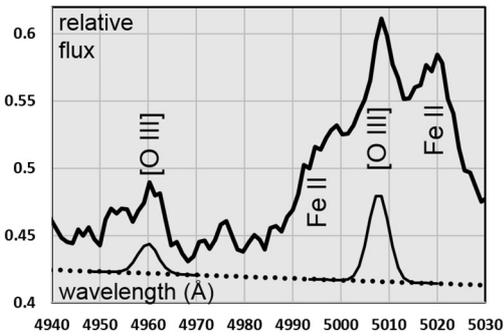

**Figure 5.** The characteristic AGN [O III] narrow lines at 4959 Å and 5007 Å barely visible amongst strong Fe II lines.

No further common Seyfert forbidden lines (e.g. [O II] 3727 Å) could be identified in the spectrum.

3.5. Further emission line characteristics
We make the following further observations about our 1H0707–495 spectrum.

No obvious changes could be confirmed between our data and previously published spectra of this object (e.g. [7]). This includes the discovery spectrum [1] measured at least 30 years before us. We also did not find any evidence of short term variability, our two sets of spectra taken six days apart in essence looking identical to each other.

With the exception of H-alpha, the spectral range observed covers the Balmer series of hydrogen. One clearly sees the emission lines from H-beta to H-epsilon in our spectrum. The relative strength of the bluest of the lines confirms that reddening due to dust in the line of sight must be low.

In addition to these and the Fe II emission we can confirm a relatively strong emission feature with peaks corresponding to the Na I doublet at $\lambda\lambda$ 5889/5896 Å. The wavelengths of other peaks always match one of the wavelengths catalogued in the extensive catalogue of emission lines identified in by Véron-Cetty et al. in I Zw 1 [8].

Even though this is not evident from figures 2 and 3, the complete spectrum of I Zw 1 (used to determine the Véron-Cetty et al. [8] Fe II template) actually matches the spectrum of 1H0707–495 presented in this paper exceptionally well (compare our figure 1 to figure 6 in that paper [8]). We mention here some minor differences, namely that the emission feature at ~3885 Å (probably Fe II]) and some emission (possibly a blend of Fe II and Fe II]) in the red wing of H-gamma.

Other than that, however, the spectra practically match each other.

## 4. Discussion and conclusion
We now compare the spectral characteristics of 1H0707–495 with those of other objects associated with the I Zw 1 class.

As discussed in the previous section, 1H0707–495 can truly be considered as a 'near-twin' to the prototype I Zw 1 itself. IRAS 07598+6508 is another relatively bright and well-studied representative of the I Zw 1 class of AGN [11], though its Fe spectrum is stronger compared to the Balmer lines, and the individual emission lines are wider.

SDSS J120011–020452 has also been shown to have exceptionally weak Seyfert forbidden lines [12], with only [O II] 3727 Å previously detected and no sign of [O III] 4959/5007 Å. There is also a strong Na I absorption feature due to interstellar material in the SDSS J120011–020452 host galaxy that we do not see at all in 1H0707–495.

The spectrum of the I Zw 1 AGN 1H0707–495 that we have presented in this paper confirm this to be an object of substantial interest, and invites further optical spectroscopy (in addition to the considerable recent studies of this AGN in x-rays). The project leading to the current paper plans the collection of a further 3-4 sets of spectra at approximately yearly intervals, to determine whether any small spectral changes can be detected, and if so which spectral lines are varying. Combining all the spectra with a total integration time amounting to 6-8 hours would reduce the noise in the spectral data sufficiently to discover hitherto undetected weak features in the spectrum and perform one of the most detailed analyses of the emission line properties in I Zw 1 objects. The comparison of this spectrum with the few high quality spectra available for similar objects of this class will hopefully enable us to gain a much better understanding of the structure of the nuclear regions in this type of AGN.

**Acknowledgements**
The paper is based on data collected with the telescopes at the South African Astronomical Observatory. We thank SAAO for the allocation of telescope time.